\documentclass[a4paper,aps,pra,reprint,superscriptaddress,nofootinbib,floatfix,longbibliography,showkeys]{revtex4-1}

\usepackage{graphicx}
\usepackage{amsmath,amssymb,amsthm,amstext,comment,mathtools,placeins}
\usepackage{natbib}
\usepackage{color,xcolor}
\usepackage{array,enumerate}
\usepackage{bm}
\usepackage{multirow}
\usepackage{booktabs}
\usepackage[breaklinks,colorlinks,citecolor=blue,linkcolor=blue]{hyperref}
\usepackage{braket}
\usepackage{txfonts}
\usepackage[normalem]{ulem}

\newcommand{\bk}{{\boldsymbol{k}}}
\newcommand{\bx}{{\boldsymbol{x}}}
\newcommand{\be}{{\boldsymbol{e}}}
\newcommand{\bE}{{\boldsymbol{E}}}
\newcommand{\bX}{{\boldsymbol{X}}}
\usepackage{booktabs}
\usepackage{siunitx}
\begin{document}
\title{Quantum-enhanced Search for Galactic Dark Photons Using Cat-State Cavities}

\author{Aoying Wang}
\thanks{These authors contributed equally to this work.}
\affiliation{School of Physics and State Key Laboratory of Nuclear Physics and Technology, Peking University, Beijing 100871, China}
\author{Yanjie Zeng}
\thanks{These authors contributed equally to this work.}
\affiliation{School of Physics and State Key Laboratory of Nuclear Physics and Technology, Peking University, Beijing 100871, China}
\affiliation{Center for High Energy Physics, Peking University, Beijing 100871, China}
\author{Yanyan Cai}
\affiliation{International Quantum Academy, Shenzhen 518048, China}
\affiliation{Southern University of Science and Technology, Shenzhen 518055, China}
\author{Jing Shu}
\email{jshu@pku.edu.cn}
\affiliation{School of Physics and State Key Laboratory of Nuclear Physics and Technology, Peking University, Beijing 100871, China}
\affiliation{Center for High Energy Physics, Peking University, Beijing 100871, China}
\affiliation{Beijing Laser Acceleration Innovation Center, Huairou, Beijing 101400, China}
\author{Bin Xu}
\email{binxu@kias.re.kr}
\affiliation{School of Physics, Korea Institute for Advanced Study, Seoul 02455, Republic of Korea}
\author{Yuan Xu}
\affiliation{International Quantum Academy, Shenzhen 518048, China}
\affiliation{Shenzhen Branch, Hefei National Laboratory, Shenzhen 518048, China}
\author{Pan Zheng}
\affiliation{International Quantum Academy, Shenzhen 518048, China}

\begin{abstract}
We present a quantum-enhanced search for the Galactic dark-photon background (GDPB), a relativistic and anisotropic vector field that can be produced by dark-matter decay or annihilation in the Milky Way halo.
We reinterpret an existing quantum-enhanced dark-photon dataset obtained with a superconducting circuit QED platform originally developed for wave-like dark-photon dark-matter searches, consisting of a high-$Q$ niobium cavity in the $\mathrm{TM}_{010}$ mode dispersively coupled to a transmon qubit.
By preparing four-component cat states and performing repeated quantum non-demolition parity measurements, analyzed with a hidden Markov model, we extract photon-transition probabilities for multiple cat amplitudes and integration times. The coupling to the GDPB is governed by a polarization- and angle-dependent cavity overlap function, which predicts a distinctive diurnal modulation of the excitation rate for both longitudinal and transverse polarizations. From the absence of such modulation, we set 90\% confidence-level bounds on the kinetic-mixing parameter $\epsilon$ for two benchmark production mechanisms (parametric and cascade decay) over the relativistic mass range $m_X \ll \omega_c$.
\end{abstract}
\maketitle

\section{Introduction}

The dark photon, a massive vector boson kinetically mixed with the standard-model photon, is a well-motivated dark matter candidate and a generic prediction of many standard model extensions, including string-inspired scenarios and hidden-sector gauge symmetries~\cite{Holdom:1985ag,svrcek2006axions,abel2008kinetic,arvanitaki2010string,goodsell2009naturally}. 
Beyond constituting a cold, non-relativistic dark-matter component~\cite{Nelson:2011sf, Arias:2012az,Graham:2015rva}, dark photons may also populate the present-day Universe as a relativistic background. 
This Galactic Dark Photon Background (GDPB) can arise from secondary emission via dark-matter decay or annihilation~\cite{dror2021cosmic, chen2022dissecting, nitta2023search, zeng2025cavity}, or from astrophysical sources such as supernovae~\cite{raffelt1999particle,eby2024diffuse,de2025star} and the collapse of boson stars~\cite{eby2016collapse,levkov2017relativistic,budker2023generic,du2024soliton}. 
Among these production channels, dark-matter decay within the Milky Way halo is of particular interest, as it traces the Galactic dark matter density profile and generates an anisotropic flux that produces a characteristic diurnal modulation at terrestrial detectors due to Earth's rotation~\cite{nitta2023search,zeng2025cavity}.

\begin{figure*}
    \centering
    \includegraphics[width=1\linewidth]{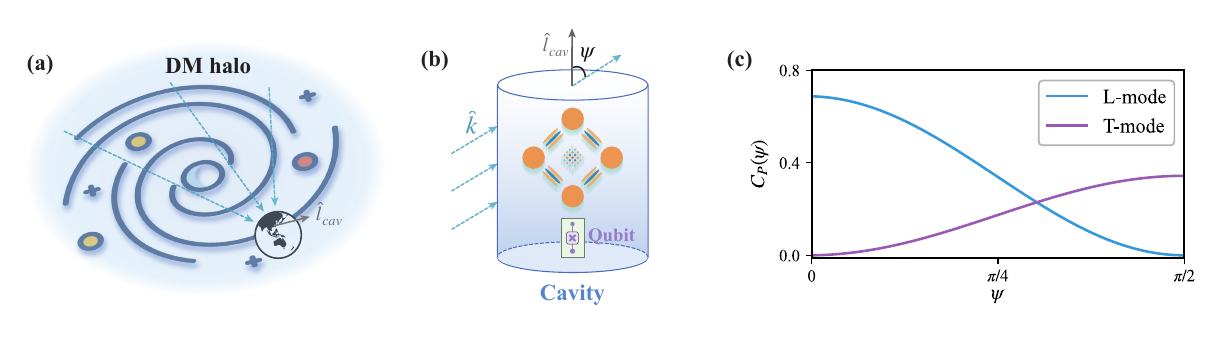}
    \caption{
\textbf{(a)} Schematic of an anisotropic GDPB produced by dark-matter decay or annihilation in the Milky Way halo. Earth's rotation changes the cavity axis $\hat{\boldsymbol{l}}_{\mathrm{cav}}$ relative to the Galactic frame and produces a diurnal modulation of the line-of-sight-integrated signal. \textbf{(b)} cQED detector comprising a transmon qubit dispersively coupled to a high-$Q$ niobium cavity initialized in a four-component cat state $|\psi_0\rangle$. A GDPB mode propagating along $\hat{\bk}$ forms an angle $\Psi$ with the cavity axis. \textbf{(c)} Cavity overlap $C_P(\Psi)$ for longitudinal (L, blue) and transverse (T, purple) polarizations. The longitudinal and transverse responses are largest at $\Psi=0$ and $\Psi=\pi/2$, respectively.}
\label{fig:fig_1}
\end{figure*}

Laboratory searches for dark photons have advanced rapidly in recent years, with superconducting \text{cav}ities emerging as particularly powerful probes owing to their extremely high quality factors \cite{Caputo:2021eaa,SHANHE:2023kxz,Zheng_2026}. 
In particular, circuit quantum electrodynamics (cQED) platforms\cite{Blais:2020qwa}, where a superconducting qubit is dispersively coupled to a high-$Q$ microwave cavity, have enabled highly sensitive  dark photon measurements through quantum non-demolition (QND) measurements of photon number, providing a route beyond the standard quantum limit (SQL) ~\cite{caves:1981,Clerk:2008nv,Sun:2014,Fang:2024ple,dixit2021searching,agrawal2024stimulated}.
Notably, nonclassical cavity states can significantly enhance the response to weak dark photon signals, especially four-component cat states which increase the signal transition rate in proportion to the cat-state size~\cite{Vlastakis:2013,Ofek:2016,Zheng_2026}.
However, these quantum sensing strategies have historically been optimized exclusively for cold, coherent dark photon dark matter (DPDM), where the signal is modeled as narrow-band and spatially uniform. They have not yet been systematically adapted to the stochastic, relativistic GDPB, which is distinguished by its broad spectral features and specific angular structure.

In this work, we extend a cat-state-based superconducting cavity--qubit platform to search for the anisotropic GDPB~\cite{Zheng_2026}. 
We construct a model for the direction- and polarization-sensitive cavity response and exploit the diurnal modulation as a key discriminator against background. 
By reinterpreting existing quantum-enhanced experimental data, we place new constraints on the kinetic-mixing parameter $\epsilon$ in the relativistic regime $m_X \ll \omega_c$, constituting the first polarization-sensitive, quantum-enhanced laboratory search for a GDPB.

The paper is organized as follows. Section~\ref{sec:background} outlines the theoretical model of GDPB. Section~\ref{sec:cavity_coupling} derives the direction- and polarization-dependent cavity response to the GDPB field. Section~\ref{sec:Response} establishes the quantum-enhanced sensing methodology. Section~\ref{sec:data} presents the experimental data analysis, including detector calibration and statistical inference. Finally, Section~\ref{sec:conclusion} concludes the paper.

\section{Galactic Dark Photon Background}
\label{sec:background}

The GDPB arises as a relativistic secondary emission from dark-matter decay or annihilation within the Milky Way halo. Unlike cold DPDM, which behaves as a coherent, non-relativistic field on laboratory timescales, the GDPB constitutes a stochastic, anisotropic vector field with distinct polarization modes. We focus on decay benchmarks and model it as a Gaussian-distributed vector field:
\begin{equation}
    \label{eq:DP_field}
    \bX(t,\bx) = \int \frac{d\omega}{2\pi} \int d^2\hat{\bk} \sum_P X_P(\omega, \hat{\bk}) \be_P(\hat{\bk}) e^{i(\omega t-\bk\cdot\bx)},
\end{equation}
where $X_P(\omega, \hat{\bk})$ is the amplitude for polarization mode $P$ (longitudinal L and transverse T) with the dispersion relation $\omega^2=|\bk|^2+m_X^2$, and $\be_P(\hat{\bk})$ denotes the corresponding polarization basis vectors in the Galactic frame. The basis $\{\be_L(\hat{\bk}), \be_{T_l}(\hat{\bk}), \be_{T_r}(\hat{\bk})\}$ spans the three physical degrees of freedom of the massive vector field. In the unitary gauge, the corresponding polarization 4-vectors are:
\begin{align}
e_L^\mu &= (k/m_X, \omega\hat{\bk}/m_X), \\
e_{T_l}^\mu &= (0,(\hat{\be}_{T_1} - i\hat{\be}_{T_2})/\sqrt{2}), \\
e_{T_r}^\mu &= (0,(\hat{\be}_{T_1} + i\hat{\be}_{T_2})/\sqrt{2}).
\end{align}
These vectors are constructed from an orthonormal spatial triad $\{\hat{\bk}, \hat{\be}_{T_1}, \hat{\be}_{T_2}\}$ tied to the propagation direction:
\begin{align}
\hat{\bk} &= (\sin\theta_k\cos\phi_k, \sin\theta_k\sin\phi_k, \cos\theta_k), \\
\hat{\be}_{T_1} &= (\cos\theta_k\cos\phi_k, \cos\theta_k\sin\phi_k, -\sin\theta_k), \\
\hat{\be}_{T_2} &= (-\sin\phi_k, \cos\phi_k, 0),
\end{align}
where $(\theta_k, \phi_k)$ are the spherical coordinates of $\hat{\bk}$.

The statistical properties of the GDPB are encoded in the two-point correlation function:
\begin{equation}
    \label{eq:correlation}
    \langle X_P(\omega,\hat{\bk})X_{P'}^*(\omega',\hat{\bk}')\rangle = \delta(\omega-\omega')\delta^{(2)}(\hat{\bk}-\hat{\bk}') \delta_{PP'}\mathcal{S}_P(\omega,\hat{\bk}).
\end{equation}
Here, $\mathcal{S}_P(\omega,\hat{\bk})$ is the polarization-dependent power spectral density (PSD). Delta functions enforce statistical independence across frequencies, directions, and polarizations. The total local energy density is normalized as
\begin{equation}
    \rho_X = \frac{1}{4\pi^2} \sum_P \int d\omega\, d^2\hat{\bk}\, \omega^2 \mathcal{S}_P(\omega,\hat{\bk}),
\end{equation}
where we adopt $\rho_X=1000\rho_\gamma$ as a representative cosmological upper bound~\cite{nitta2023search,dror2021cosmic,zeng2025cavity}, with $\rho_\gamma \approx 0.26\,\mathrm{eV/cm}^3$ being the local cosmic microwave background energy density~\cite{yao2006review,Fixsen:2009ug}.

The PSD $\mathcal{S}_P(\omega,\hat{\bk})$ encodes three physical ingredients including the underlying interaction, decay kinematics, and Galactic dark matter profile.
The interaction type determines the polarization composition: a dark Higgs-like coupling ($\Phi X^\mu X_\mu$) generates longitudinal modes, while an axion-like coupling ($\Phi X_{\mu\nu}\tilde{X}^{\mu\nu}$) yields transverse modes~\cite{zeng2025cavity}. In the unitary gauge, the spatial polarization vectors satisfy $|\be_T|^2 = 1$ for transverse modes and $|\be_L|^2 = \omega^2/m_X^2$ for the longitudinal mode. In the relativistic limit ($\omega \gg m_X$), this geometric factor enhances the longitudinal response compared to non-relativistic DPDM, where all polarizations contribute comparably at a fixed energy density.

The spectral shape is determined by the decay kinematics.
We consider two representative production mechanisms~\cite{zeng2025cavity}: 
(i) two-body parametric decay ($\Phi \to XX$), producing a narrow spectral line characterized by a quality factor $Q \simeq 500$;
(ii) four-body cascade decay ($\chi \to \Phi\Phi \to XXXX$), yielding a broad continuum with a kinematic cutoff.
The corresponding PSDs are:
\begin{align}
    \label{eq:S_PD}
    \mathcal S_P^{\rm PD}(\omega,\hat{\bk}) &= \frac{4\pi Q\rho_X}{\omega m_{\rm DM}^2\bar J}J(\theta_k) \Theta\!\left(\frac{m_{\rm DM}}{4Q}-\left|\omega-\frac{m_{\rm DM}}{2}\right|\right),\\
    \label{eq:S_CD}
    \mathcal S_P^{\rm CD}(\omega,\hat{\bk}) &= \frac{8\pi\rho_X}{\omega m_{\rm DM}^2\bar J}J(\theta_k) \Theta\!\left(\frac{m_{\rm DM}}{2}-\omega\right).
\end{align}

Both decay channels share an angular distribution determined by the Galactic $J$-factor, which represents the line-of-sight integral of the dark matter halo density:
\begin{equation}
    \label{eq:J_factor}
    J(\theta_k) = \frac{1}{r_{\odot}} \int_{\rm l.o.s.} \left(\frac{\rho^{\mathrm{DM}}(r(s,\theta_k))}{\rho_s}\right) ds,
\end{equation}
where $r(s,\theta_k) = (r_{\odot}^2 + s^2 - 2 r_{\odot} s \cos\theta_k)^{1/2}$ is the Galactocentric distance, with $s$ denoting the line-of-sight distance from Earth and $\theta_k$ the angle between the observation direction $\hat{k}$ and the Galactic Center. Assuming a standard Navarro, Frenk, and White (NFW) profile~\cite{cirelli2011pppc} with scale radius $r_s = 24.42\,\mathrm{kpc}$ and characteristic density $\rho_s = 0.184\,\mathrm{GeV/cm}^3$~\cite{Navarro:1996gj}, alongside Earth's Galactocentric distance $r_{\odot} = 8.33\,\mathrm{kpc}$, the sky-averaged normalization yields $\bar{J} \approx 2.19$. As Earth rotates, this spatial anisotropy, combined with the directional cavity response, produces a diurnal modulation of the detected signal for both decay models, as we show in Sec.~\ref{sec:cavity_coupling}.

\section{directional cavity response to GDPB}
\label{sec:cavity_coupling}

To bridge the GDPB signal model $\mathcal{S}_P(\omega, \hat{\bk})$ to the detector response, we consider the interaction Hamiltonian between the stochastic vector field and the cavity electromagnetic field, $H_{\text{int}} = \int_V \bm{A} \cdot \boldsymbol{J}_{\text{eff}} \, dV$. The background field induces an effective current $\boldsymbol{J}_{\text{eff}} = \epsilon m_X^2 \bX(t,\bx)$, where $\epsilon$ is the kinetic mixing coefficient. Restricting the quantized cavity field to its fundamental $\mathrm{TM}_{010}$ mode (with resonant frequency $\omega_c$ and electric-field profile $\mathbf E_0(\mathbf x)$), the coupling to an incoming dark-photon mode is determined by the spatial overlap between the dark-photon field and the cavity mode:
\begin{equation}
    C_P(\hat{\bk}) \equiv \frac{1}{V} \left| \int_V \hat{\be}_P(\hat{\bk}) \cdot \bE_0(\bx) e^{-i\bk\cdot\bx} dV \right|^2,
\end{equation}
where $V$ is the cavity volume and $\hat{\be}_P$ is the unit polarization vector. 

Unlike non-relativistic DPDM ($|\bk|\ll m_X$), where the spatial phase factor $e^{-i\bk\cdot\bx}\simeq 1$ reduces the overlap to a polarization-weighted volume integral~\cite{Zheng_2026}, the relativistic GDPB ($|\bk|\simeq\omega_c$) preserves the full $e^{-i\bk\cdot\bx}$ term. This introduces significant direction-dependent interference within the cavity. 

For the axisymmetric $\mathrm{TM}_{010}$ mode, the overlap integral depends strictly on the incidence angle $\Psi(t)$ between the dark-photon arrival direction $-\hat{\bk}$ and the instantaneous cavity symmetry axis $\hat{\bm{l}}_{\text{cav}}(t)$ (see Fig.~\ref{fig:fig_1}):
\begin{equation}
    \Psi(t) \equiv \arccos\left( \hat{\bk}\cdot\hat{\bm{l}}_{\text{cav}}(t)\right),
\end{equation}
The overlap factor thus simplifies to $C_P[\Psi(t)]$.

To evaluate the temporal evolution of $\Psi(t)$, we account for the rotation of the terrestrial laboratory relative to the Galactic frame. 
Unlike previous searches that assume a vertically aligned cavity axis~\cite{zeng2025cavity}, we formulate a generalized treatment for arbitrary orientations applicable to future directional sensor networks.
For a generic cavity orientation~\cite{zeng2025cavity} at latitude $\varphi_{\mathrm{cav}}$ and longitude $\lambda_{\mathrm{cav}}$, the axis in the local East--North--Up (ENU) frame is defined by azimuth $A$ (eastward from North) and zenith angle $B$:
\begin{equation}
    \hat{\bm{l}}_{\text{cav}} = \sin B \left(\cos A \cdot \hat{\bm{N}} +  \sin A \cdot \hat{\bm{E}} \right) + \cos B \cdot \hat{\bm{U}},
\end{equation}
where $\hat{\bm{E}}$, $\hat{\bm{N}}$, and $\hat{\bm{U}}$ are the local unit vectors:
\begin{align}
    \hat{\bm N}
&=
\left(
-\sin\varphi_{\rm cav}\cos\lambda_{\rm cav},
-\sin\varphi_{\rm cav}\sin\lambda_{\rm cav},
\cos\varphi_{\rm cav}
\right),\\
\hat{\bm E}
&=
\left(
-\sin\lambda_{\rm cav},
\cos\lambda_{\rm cav},
0
\right),\\
\hat{\bm U}
&=
\left(
\cos\varphi_{\rm cav}\cos\lambda_{\rm cav},
\cos\varphi_{\rm cav}\sin\lambda_{\rm cav},
\sin\varphi_{\rm cav}
\right).
\end{align}
Earth's rotation transforms this stationary terrestrial vector into a dynamic one within the Earth-centered Galactic Center (GC) frame, where the $-\hat{\bm z}_g$ axis aligns with the line of sight to the GC. In this frame, the cavity axis is $\hat{\bm l}_{\mathrm{cav}}(t) = (\sin\theta_{\mathrm{cav}}(t)\cos\phi_{\mathrm{cav}}(t), \sin\theta_{\mathrm{cav}}(t)\sin\phi_{\mathrm{cav}}(t), \cos\theta_{\mathrm{cav}}(t))$. 

The explicit coordinate transformation yields the time-dependent polar angle relative to the GC:
\begin{equation}
\begin{aligned}
    \cos\theta_{\text{cav}}(t) =& \sin B \Big\{ \cos A \left[\cos\delta_{\text{GC}} \sin\varphi_{\text{cav}} \cos H(t) -\sin\delta_{\text{GC}} \cos\varphi_{\text{cav}} \right] \\
    &+ \sin A \cos\delta_{\text{GC}} \sin H(t) \Big\} \\ 
    &- \cos B\left(\cos \delta_{\text{GC}}\cos \varphi_{\text{cav}}\cos H(t) + \sin \delta_{\text{GC}}\sin \varphi_{\text{cav}}\right),
\end{aligned}
\end{equation}
where $(\alpha_{\text{GC}}, \delta_{\text{GC}})$ are the equatorial coordinates of the GC, and $H(t) = \frac{2\pi t}{24} + \lambda_{\text{cav}} - \alpha_{\text{GC}}$ is its local hour angle. For our horizontal East--West setup ($A=90^\circ$, $B=90^\circ$), this reduces to $\cos\theta_{\text{cav}}(t) = \cos\delta_{\text{GC}}\sin H(t)$.
The resulting evolution of $\hat{\bm l}_{\mathrm{cav}}(t)$ modulates the directional overlap $C_P[\Psi(t)]$. 
For the axisymmetric $\mathrm{TM}_{010}$ mode and a GDPB distribution symmetric about the GC axis, the azimuthal integration reduces the response to a function of $\theta_{\mathrm{cav}}(t)$ alone, yielding distinct diurnal templates for longitudinal and transverse polarizations.

\section{Quantum-enhanced cavity sensing}
\label{sec:Response}

Having defined the stochastic GDPB through $\mathcal{S}_P(\omega, \hat{\bk})$ and the directional overlap $C_P(\Psi)$, we describe how the resulting weak cavity displacement is amplified and read out using cat-state quantum sensing. 
We employ an established cQED platform~\cite{Zheng_2026} which uses repeated QND parity measurements for cavity-state readout~\cite{Sun:2014}. 
The system comprises a high-$Q$ niobium cavity ($\omega_c=2\pi \times 6.442$\,GHz, $Q \sim 1.86\times10^8$) dispersively coupled to a transmon qubit ($\omega_q= 2\pi \times 5.205 $\,GHz, $T_1 = 175\,\mu\mathrm{s}$) serving as the ancilla for state preparation and readout.

Quantum enhancement is achieved by preparing the cavity in a four-component cat (compass) state. The orthogonal compass basis is:
\begin{align}
|\phi_0\rangle&=N_{4,0}(|\alpha\rangle+|-\alpha\rangle+|i\alpha\rangle+|-i\alpha\rangle),\\
|\phi_1\rangle&=N_{4,1}(|\alpha\rangle-|-\alpha\rangle-i|i\alpha\rangle+i|-i\alpha\rangle),\\
|\phi_2\rangle&=N_{4,2}(|\alpha\rangle+|-\alpha\rangle-|i\alpha\rangle-|-i\alpha\rangle),\\
|\phi_3\rangle&=N_{4,3}(|\alpha\rangle-|-\alpha\rangle+i|i\alpha\rangle-i|-i\alpha\rangle).
\end{align}
This encoding allows us to distinguish the GDPB signal from the dominant single-photon-loss background. Starting from $|\phi_0(\alpha)\rangle$, a weak GDPB-induced displacement can drive transitions to both $|\phi_1(\alpha)\rangle$ and $|\phi_3(\alpha)\rangle$, whereas single-photon loss primarily induces transitions to $|\phi_3(\alpha)\rangle$. We therefore choose transitions to $|\phi_1(\alpha)\rangle$ as the signal indicator to suppress contamination from photon loss.

Under the stochastic dark photon field, the cavity evolves via $U(\tau) = \mathcal{T}\exp\left(-i\int_0^\tau H_{\text{int}}(t) dt\right) e^{-i\omega_c\hat{a}^\dagger\hat{a}\tau}$. Interaction with a single momentum mode $k$ induces an effective displacement $D(z_k)|\phi_0\rangle$ with amplitude:
\begin{equation}
z_k = -\frac{\epsilon m_X^2}{\sqrt{2\omega_c}} \sum_P X_P(\omega, \hat{\bk}) \left[C_P(\hat{\bk}) V\right]^{1/2} \frac{1 - e^{-i\delta\tau}}{\delta},
\end{equation}
with $\delta \equiv \omega - \omega_c$ denoting the detuning.
To first order in weak driving ($|z_k|\ll1$), the transition amplitude is $c_{1,k} \simeq \langle\phi_1|D(z_k)|\phi_0\rangle \approx \alpha^* z_k$. Considering the complete dark photon field encompassing all momentum modes, the collective signal amplitude accumulates as $c_1 \approx \int \frac{d\omega}{2\pi} d^2\hat{\bk}\, c_{1,k}$. The transition probability is obtained through ensemble averaging over the stochastic field, $P_{\alpha}(\tau) = \langle|c_1|^2\rangle$
, yielding:
\begin{equation}
\begin{aligned}
    \label{eq:P_integral}
    P_{\alpha}(\tau) = &\frac{\epsilon^2 m_X^4 V |\alpha|^2}{2\omega_c}\sum_{P} |\be_P|^2 \\ &\int \frac{d\omega}{4\pi^2} d^2\hat{\bk} \left[\frac{\sin(\delta\tau/2)}{\delta/2}\right]^2 \mathcal{S}_P(\omega, \hat{\bk}) C_P(\hat{\bk}).\\
\end{aligned}
\end{equation}


Evaluating this spectral integral requires comparing the integration time $\tau$ to the intrinsic coherence time of the GDPB, $\tau_{\mathrm{GDPB}} \sim 1/\Delta\omega$, which is determined by the dark photon production mechanism:
\begin{equation}
\tau_{\mathrm{GDPB}} \sim \frac{1}{\Delta\omega} = 
\begin{cases}
2Q/m_{\mathrm{DM}} & \text{parametric decay,}\\
2/m_{\mathrm{DM}} & \text{cascade decay.}
\end{cases}
\end{equation}
For $m_X \sim 10^{-5}$ eV, the trial integration time ($\tau \sim \mathcal{O}(\mathrm{ms})$) heavily exceeds the background coherence time ($\tau_{\rm GDPB}\sim\mathcal O(10\,\mathrm{ns})$ for parametric decay and $\mathcal O(10\,\mathrm{ps})$ for cascade decay). In this long-time regime ($\tau \gg \tau_{\mathrm{GDPB}}$), the finite-time spectral kernel in Eq.~\eqref{eq:P_integral} reduces to the asymptotic limit $[\sin(\delta\tau/2)/(\delta/2)]^2 \approx 2\pi \tau \delta(\omega-\omega_c)$.

Applying this limit, statistically independent field components accumulate incoherently, and the transition probability simplifies to a linear growth with $\tau$:
\begin{equation}
\label{eq:probability}
P_{\alpha}(t, \tau) = \frac{\epsilon^2 m^4_{X}V |\alpha|^2\tau }{4\pi\omega_c} \sum_P |\be_P|^2 \int \mathcal{S}_P\left(\omega_c, \hat{\bk}\right) C_P\left(\Psi(t)\right) d^2\hat{\bk},
\end{equation}
where $t$ is the laboratory time fixing the cavity orientation relative to the Galactic frame. 
This $P_\alpha \propto \tau$ dependence provides a distinct temporal signature for the short-coherence GDPB, contrasting sharply with the $P \propto \tau^2$ scaling characteristic of coherent dark photon dark matter (DPDM) when $\tau \ll \tau_{\mathrm{DM}}$.

Despite the incoherent temporal accumulation, the cat state retains a quantum advantage. The amplitude $\alpha$ coherently enhances the displacement sensitivity by coupling $|\phi_0\rangle$ to $|\phi_1\rangle$, boosting the excitation probability by a factor of $|\alpha|^2$ relative to a vacuum measurement. This extends the quantum-sensing advantage demonstrated for coherent DPDM~\cite{Zheng_2026} to the stochastic GDPB regime.

Finally, the angular integral in Eq.~\eqref{eq:probability} combines the Galactic sky distribution with the instantaneous cavity response. The power spectral density $\mathcal{S}_P(\omega_c,\hat{\bk})$ provides the spectral-angular weight, while $C_P(\Psi(t))$ dictates the cavity coupling efficiency. As Earth rotates, the changing cavity orientation imprints the deterministic diurnal modulation derived in Sec.~\ref{sec:background}.

\section{Data analysis and results}
\label{sec:data}

\begin{figure*}[t]
    \centering
    \includegraphics[width=0.92\textwidth]{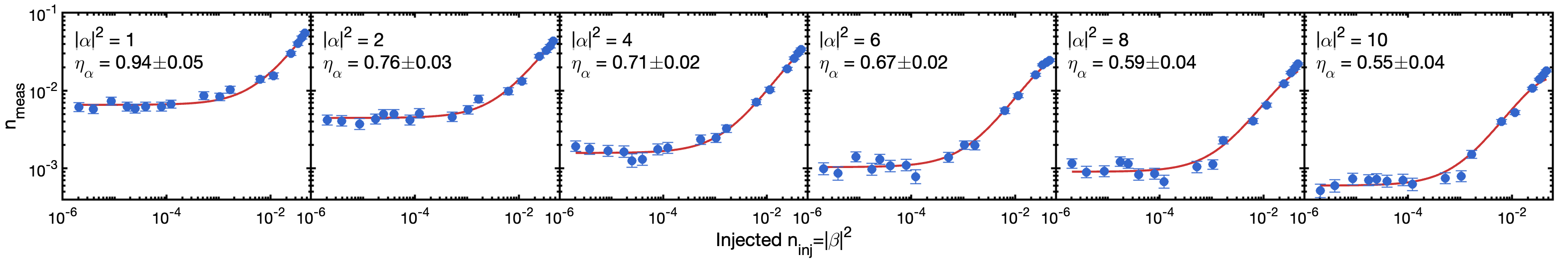}
    \caption{Detector calibration: measured positive-event fraction $n_{\mathrm{meas}}$ versus injected photon number $n_{\mathrm{inj}}=|\beta|^2$ for the six cat-state sizes. Solid curves are the linear fits $n_{\mathrm{meas}}=\eta_\alpha|\alpha|^2n_{\mathrm{inj}}+\delta_\alpha$ used to determine the efficiency and false-positive probability.}
    \label{fig:eta_fit}
\end{figure*}

In this section, we reinterpret an existing quantum-enhanced dark-photon dataset to search for the anisotropic GDPB. 
Although the original experiment targeted coherent DPDM, the primary observable, the photon-transition probabilities extracted from parity-resolved readout remain sensitive to a relativistic stochastic flux when analyzed in the time domain. 
We exploit the characteristic diurnal modulation of the GDPB, induced by Earth's rotation relative to the Galactic frame and the directional cavity overlap. 
The measured excitation rates are therefore correlated with laboratory time and compared with theoretical modulation templates for longitudinal and transverse polarizations.

Our quantum measurements were performed in September 2024 in Shenzhen, China ($22.55^\circ$N, $114.05^\circ$E), with the cavity symmetry axis oriented horizontally along the East-West direction. 
The dataset contains calibration (September 12--14) and a search (September 16--17) runs, both performed with six cat states ($|\alpha|^2=1,2,4,6,8,10$). 
For each configuration, defined by the initial state $|\phi_0(\alpha)\rangle$ and either an injected displacement amplitude (calibration) or an integration time $\tau$ (search), we acquired $N_\text{trials} \sim 10,000$ independent sequences. 
Each trial consisted of (i) preparing the cavity in $|\phi_0(\alpha)\rangle$, 
(ii) applying a controlled displacement $\hat D(\beta)$ (calibration) or free evolution for $\tau$ (search), and 
(iii) performing $M=29$ repeated QND parity measurements. 
The resulting binary qubit-readout sequences are analyzed using a hidden Markov model (HMM)~\cite{Rabiner:1989}, 
reconstructing cavity-state posterior probabilities  while accounting for measurement errors and state fluctuations. 
The reconstructed posterior translates physical state evolution into a transition decision while rigorously retaining readout uncertainty. 
For each trajectory, HMM gives the posterior state probability $P(\phi_j)$, from which we compute the likelihood ratio $\lambda=P(\phi_1)/[P(\phi_0)+P(\phi_2)+P(\phi_3)]$. Thresholding $\lambda>\lambda_{\rm thresh}=84$ classifies the trajectory as a signal event. The total count of positive events $N_{\rm meas}$ across $N_{\rm trials}$ yields the measured detection probability $n_{\rm meas}=N_{\rm meas}/N_{\rm trials}$, which serves as the fundamental observable for both detector calibration and the modulation search.

\subsection{Detector Characterization and Calibration Procedure}
\label{sec:calibration}

Prior to the dark photon search, a comprehensive detector calibration is performed to quantify its response to simulated signals, establishing the relationship between a known physical interaction and the classified outcome $n_{\rm meas}$. Following initialization, we inject a calibrated coherent signal $\hat{D}(\beta)$ that mimics a weak dark photon field, quantified by the injected photon number $n_{\mathrm{inj}} = |\beta|^2 \ll 1$.

The detector response manifests as a linear relationship between the classified event fraction and the injected excitation:
\begin{equation}
    n_{\mathrm{meas}} = \eta_\alpha |\alpha|^2 n_{\mathrm{inj}} + \delta_\alpha.
\end{equation}
Linear regression determines two critical parameters: the detection efficiency $\eta_\alpha$ and the zero-signal false-positive intercept $\delta_\alpha$. The explicit slope factor $|\alpha|^2$ confirms the quantum enhancement provided by the cat-state encoding. The efficiency $\eta_\alpha$ corrects for information loss between a true physical excitation and algorithmic classification, whereas $\delta_\alpha$ isolates events generated by technical noise or background relaxation in the absence of an external displacement.

\begin{table}[b]
\centering
\caption{Calibrated detection efficiency $\eta_\alpha$, false-positive probability $\delta_\alpha$, and search-trial counts $N_{\rm trials}$ for each cat-state size.}
\label{tab:parameters}
\begin{tabular}{c S[table-format=1.3] S[table-format=1.2e-3] S[table-format=4]}
\hline
$|\alpha|^2$ & {$\eta_\alpha$} & {$\delta_\alpha$} & {$N_{\rm trials}$}\\
\hline
1  & 0.94 & 6.57e-3 & 9350\\
2  & 0.76 & 4.46e-3 & 5439\\
4  & 0.71 & 1.58e-3 & 5338\\
6  & 0.67 & 1.04e-3 & 5448\\
8  & 0.59 & 9.06e-4 & 5340\\
10 & 0.55 & 6.01e-4 & 5077\\
\hline
\end{tabular}
\end{table}

As shown in Fig.~\ref{fig:eta_fit} and detailed in Table~\ref{tab:parameters}, the response remains linear throughout the calibration range. The extracted efficiency drops from $0.94$ at $|\alpha|^2=1$ to $0.55$ at $|\alpha|^2=10$, while the false-positive probability concurrently decreases by an order of magnitude. Strict false-positive subtraction is imperative; an uncorrected intercept would masquerade as a spurious, cat-size-dependent physical excitation rate during the search.

\subsection{Data Reinterpretation and Modulation Search}
\label{sec:Search}

To search for the GDPB, we analyze the dependence of the measured transition probability $n_{\mathrm{meas}}$ on both the integration time $\tau$ and laboratory time $t$. 
For each experimental configuration $i$ characterized by a cat-state amplitude $\alpha_i$ and integration time $\tau_i$, we define the false-positive-corrected normalized excitation rate:
\begin{equation}
R_i = \frac{n_{\text{meas},i}-\delta_{\alpha_i}}{\eta_{\alpha_i} |\alpha_i|^2 \tau_i }.
\end{equation}
The numerator removes the static detector background, while the denominator accounts for algorithmic efficiency, quantum enhancement, and integration time. Thus, $R_i$ estimates the physical cavity-excitation rate.

\begin{figure}[t]
\centering
\includegraphics[width=0.82\linewidth]{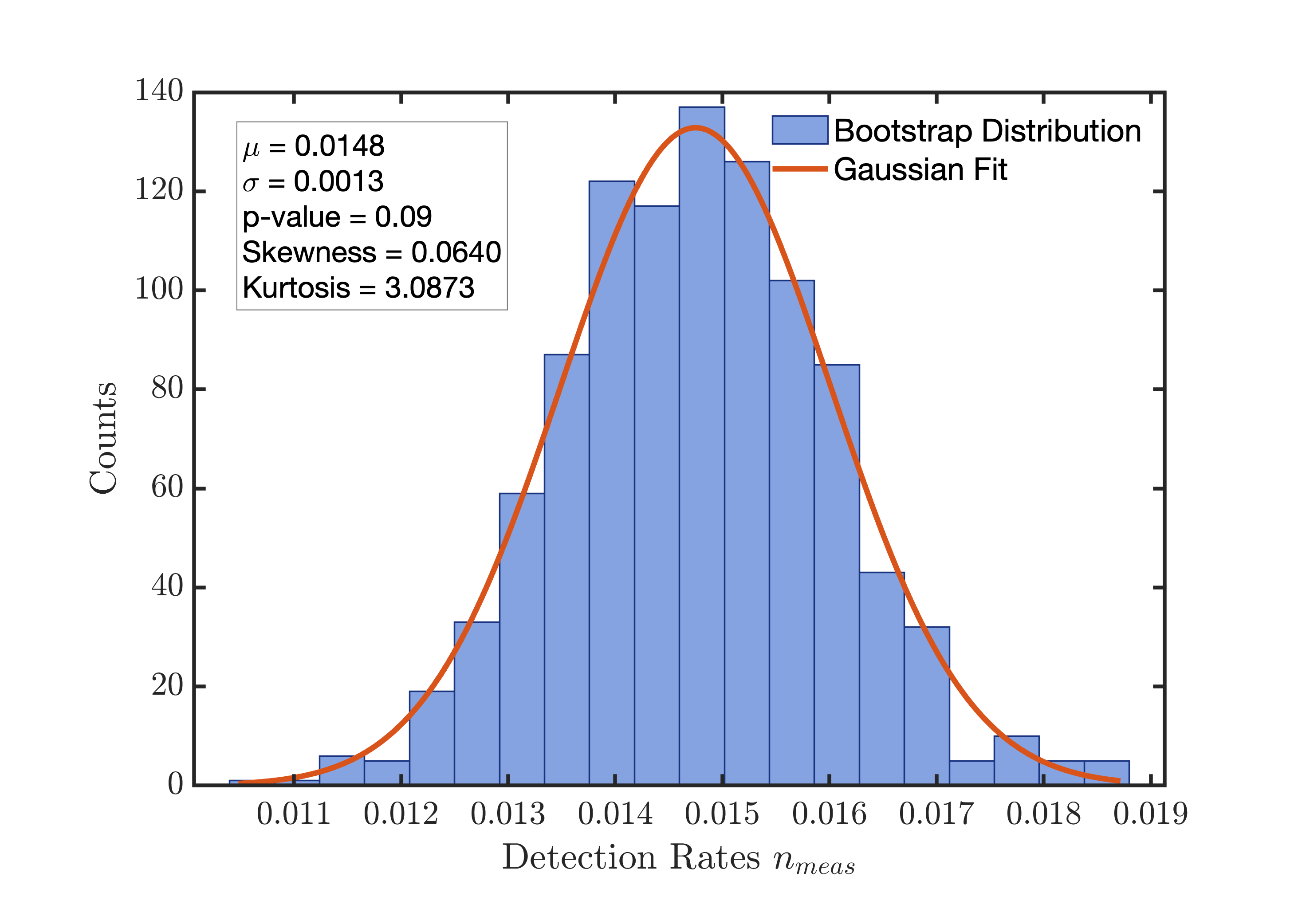}
\caption{Bootstrap distribution of $n_{\mathrm{meas}}$ for a representative search configuration ($|\alpha|^2=1, \tau=4.2$ ms). The blue histogram represents 1000 bootstrap resamples; the red curve is a Gaussian fit ($\mu \simeq 0.0148$, $\sigma \simeq 1.3\times10^{-3}$). The Shapiro--Wilk normality test yields $p \simeq 0.09$, indicating no statistically significant deviation from a normal distribution at 95\% confidence.}
\label{fig:bootstrap}
\end{figure}

The statistical fluctuations of $n_{\text{meas},i}$ are fundamentally governed by a binomial process. Given the large number of trials ($N_{\rm trials} \sim 10^4$), the central limit theorem implies these counting statistics closely approximate a Gaussian distribution. We validate this directly via bootstrap resampling of the binary outcomes (Fig.~\ref{fig:bootstrap}). For a representative configuration, the Shapiro--Wilk test yields $p \simeq 0.09$, a skewness of $0.06$, and a kurtosis of $3.09$. Because $n_{\text{meas},i}$ is normally distributed, and the mapping to $R_i$ is affine, $R_i$ also follows a Gaussian distribution.

To isolate the modulated GDPB signal from any residual static offsets, we compute the trial-count-weighted mean $\bar R$ and normalize the residual against the theoretical reference template. The signal estimator for the kinetic mixing squared is
\begin{equation}
    p_i = \frac{R_i-\bar{R}}{R_{\text{ref},i}-\bar{R}_{\text{ref}}},
\end{equation}
where $R_{\mathrm{ref},i}=P_\alpha(t_i,\tau_i,\epsilon=1)/(|\alpha_i|^2\tau_i)$ is the predicted physical rate for unit kinetic mixing. Both the data and the reference template are centered using the same statistical weights. Note that the detector false-positive probability $\delta_{\alpha_i}$ is subtracted strictly from the measured fraction, never from the physical reference $R_{\mathrm{ref},i}$.

\begin{table}[b]
\centering
\caption{Experimental parameters and associated systematic uncertainties propagated in the statistical analysis.}
\begin{tabular}{lcc}
\hline
Parameter $\Theta$ & Value &  $\sigma_{\Theta}$ \\
\hline
$\omega_c/(2\pi)$ & 6.442 GHz & 9.4 Hz \\
$ V $ & $ 19.34 \text{ cm}^3$&  $0.17 \text{ cm}^3$ \\
$C_P(\hat{k})$ &$ 0 \sim 0.7$ & $1\%$ (relative) \\
$\delta_\alpha$ & $\sim 10^{-3}$ & $\sim 10^{-4}$\\
$ \eta_\alpha $ & $ 0.55 \sim 0.94$&  $\sim 0.02$ \\
$ |\alpha|^2 $ & $ 1 \sim 10 $&  $\sim 0.03$ \\
$ \tau $ & $  0.26 \sim 4.20 \text{ ms}$& negligible \\
\hline
\end{tabular}
\label{tab:uncertainties}
\end{table}

\begin{figure}[t]
	\centering
	\includegraphics[width=\linewidth]{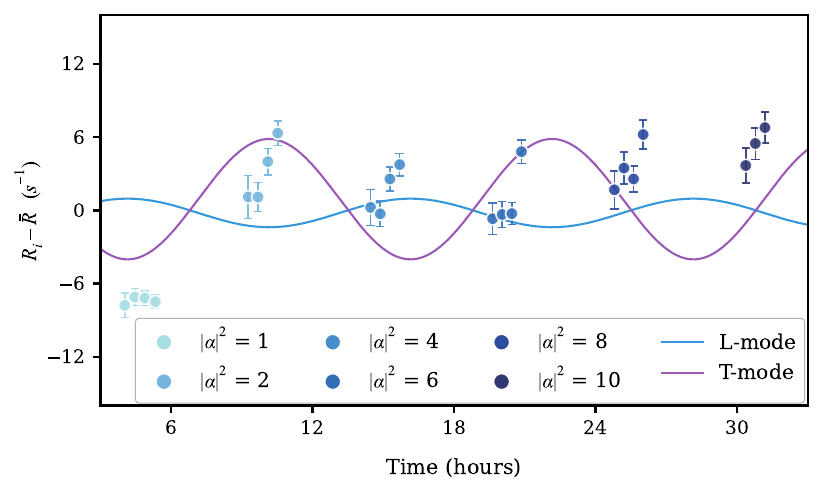}
    \caption{False-positive-corrected excitation-rate residuals $R_i-\bar R$ as a function of laboratory time. Data points become darker with increasing $|\alpha|^2$. Solid curves represent the longitudinal (blue) and transverse (purple) diurnal templates evaluated at the 90\% credible upper limit, $\epsilon_{90\%}^2R_{\mathrm{ref}}$. $t>24$ h denotes September 17.}
	\label{fig:ni_LT}
\end{figure}

Fig.~\ref{fig:ni_LT} shows the background-subtracted excitation rates $R_i-\bar{R}$ for all configurations. 
The corresponding $90\%$ C.L. upper-limit templates are shown as $\epsilon_{90\%}^2R_{\mathrm{ref}}$. 
We overlay the predicted diurnal modulations for longitudinal and transverse polarizations, whose time dependence is determined by the cavity overlap function $C_P(\Psi)$ and Galactic geometry. 
No statistically significant correlation with either template is observed.

Given the absence of a statistically significant modulation, we constrain the kinetic-mixing parameter $\epsilon$ by comparing the measured residuals with the theoretical modulation template.
For a fixed dark-photon mass $m_X$, the signal phase and temporal structure are determined by the Galactic geometry and cavity response, leaving only an overall amplitude proportional to
$\epsilon^2$.
For each configuration, the event counts follow binomial statistics. 
In the large-sample limit ($N_{\rm trials}\sim10^4$), the central limit theorem motivates the Gaussian approximation
$p_i\sim\mathcal{N}(\epsilon^2,\sigma_{p_i}^2)$, 
which is validated by bootstrap resampling.
We therefore construct
\begin{equation}\label{eq:likelihood}
    \mathcal{L}(\{p_i\}|\epsilon,m_X) \propto \prod_i \exp \left(
-\frac{(p_i-\epsilon^2)^2}{2\sigma_{p_i}^2}
\right).
\end{equation}
The total uncertainty $\sigma_{p_i}$ is derived by propagating statistical fluctuations and all systematic calibration uncertainties (summarized in Table~\ref{tab:uncertainties}) through the linear transformations:
\begin{equation}
\sigma_{p_i}^2=
\frac{\sigma_{R_i}^2+\sigma_{\bar R}^2}
{(R_{\mathrm{ref},i}-\bar R_{\mathrm{ref}})^2}
+p_i^2\!\left[
4\frac{\sigma_{\omega_c}^2}{\omega_c^2}
+\frac{\sigma_{C_P}^2}{C_P^2}
+\frac{\sigma_V^2}{V^2}\right],
\end{equation}
where the variance of the corrected rate is
\begin{equation}
\sigma_{R_i}^2=\sigma_{R,\mathrm{stat}}^2+
\frac{\sigma_\delta^2}{\eta_{\alpha_i}^2|\alpha_i|^4\tau_i^2}
+ R_i^2\left(\frac{\sigma_\eta^2}{\eta_{\alpha_i}^2}+
\frac{\sigma_{|\alpha|^2}^2}{|\alpha_i|^4}+
\frac{\sigma_{\tau_i}^2}{\tau_i^2}\right),
\end{equation}
and $\sigma_{\bar R}^2=\sigma_{R,\mathrm{stat}}^2/N_{\mathrm{total}}$. Here, $\sigma_{|\alpha|^2}$ is conservatively estimated from preparation infidelity $(1-F)$. This error model distinguishes uncertainties associated with the experimentally inferred excitation rates ($\sigma_\eta$, $\sigma_\delta$, $\sigma_{|\alpha|^2}$, $\sigma_{\tau_i}$) from those associated with the reference signal prediction ($\sigma_{\omega_c}$, $\sigma_{\mathcal C^P}$, $\sigma_V$). The former arise from detector calibration, state preparation, and timing, while the latter originate from cavity characterization and modeling of the directional response.

\begin{figure}[t]
    \centering
    \includegraphics[width=\linewidth]{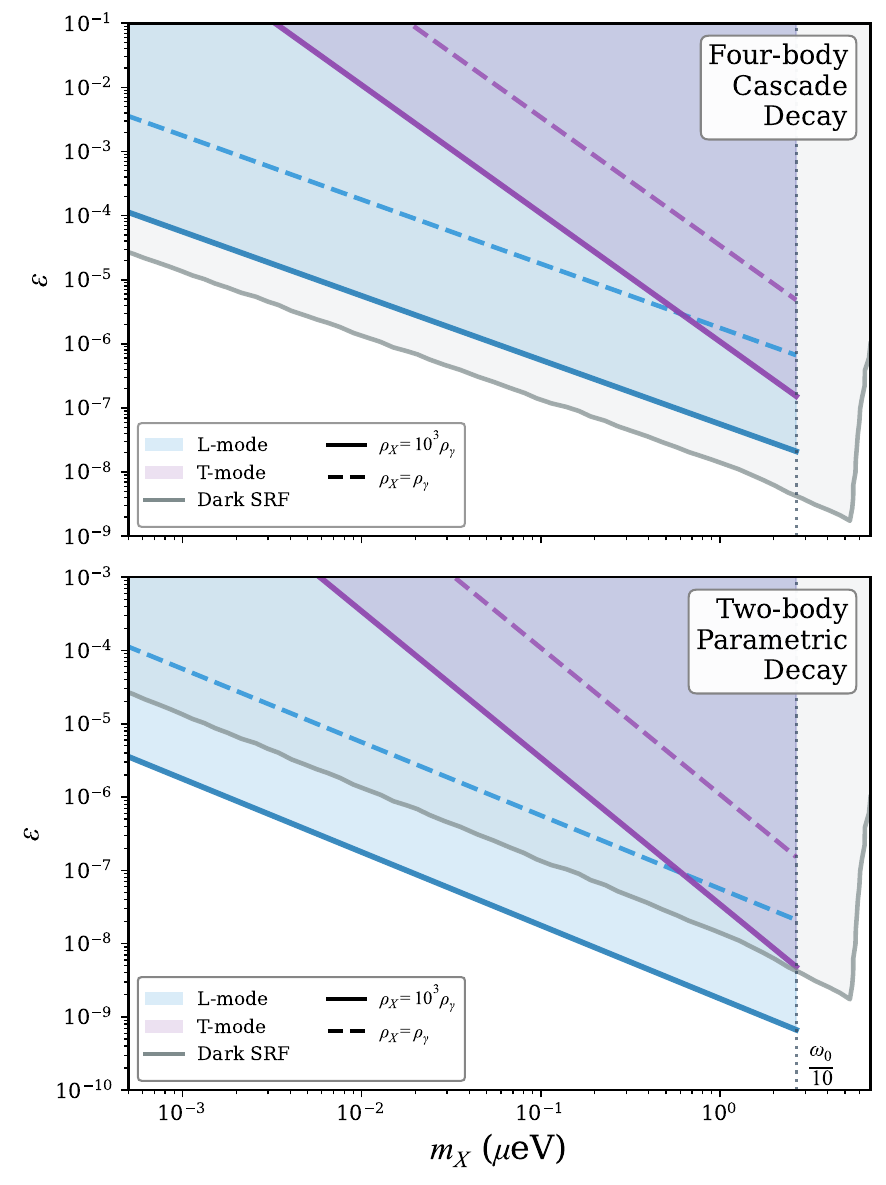}
    \caption{90\% credible upper limits on the kinetic mixing $\epsilon$ versus dark photon mass $m_X$. The upper and lower panels show four-body cascade decay and two-body parametric decay models, respectively. Solid and dashed curves correspond to local dark photon densities $\rho_X=10^3\rho_\gamma$ and $\rho_X=\rho_\gamma$. The dotted vertical line denotes $m_X=\omega_c/10$, the bound for the relativistic approximation. Gray shaded areas indicate existing Dark SRF limits~\cite{romanenko2023search}.}
    \label{fig:limits}
\end{figure}

Since no statistically significant modulation is observed, we derive Bayesian upper limits on the kinetic-mixing parameter $\epsilon$. Adopting a uniform prior in $\epsilon^2 \geq 0$, the posterior distribution is proportional to the likelihood in Eq.~\eqref{eq:likelihood}. The 90\% credible upper limit $\epsilon_{90\%}$ is then determined by
\begin{equation}
\frac{\displaystyle\int_0^{\epsilon_{90\%}^2} \mathcal L(\{p_i\}|\epsilon,m_X)\,d\epsilon^2}
{\displaystyle\int_0^\infty \mathcal L(\{p_i\}|\epsilon,m_X)\,d\epsilon^2}=0.90.
\end{equation}
The resulting 90\% credible upper limits are shown in Fig.~\ref{fig:limits}. 
We consider two benchmark production mechanisms from scalar dark matter decay: two-body parametric decay (yielding a narrow spectral line, $Q\simeq500$) and four-body cascade decay (yielding a broad continuum), both normalized to a reference energy density $\rho_X = 1000\,\rho_\gamma$. 
Our analysis is restricted to the relativistic regime $m_X \leq \omega_c/10$, where polarization purity is well-defined and diurnal modulation remains a robust discriminator.
In this regime, longitudinal modes yield significantly stronger constraints due to the relativistic enhancement factor $|\be_L|^2 = \omega_c^2/m_X^2$, whereas transverse modes are limited by $|\be_T|^2 = 1$.

\section{Conclusion}
\label{sec:conclusion}

We have applied a time-domain modulation analysis to photon-transition data from a cat-state cQED detector. Six cat-state sizes were analyzed using repeated QND parity measurements and HMM reconstruction. Coherent-displacement data supplied the detection efficiency and false-positive probability needed to place the different configurations on a common corrected-rate scale.

The GDPB model was used to generate longitudinal and transverse templates that include the polarization-resolved PSD, the Galactic $J$-factor, the laboratory coordinates, the cavity orientation, and the full directional overlap of the $\mathrm{TM}_{010}$ mode. After removal of the weighted time-independent component, the measured residuals showed no significant correlation with either template. The likelihood analysis therefore gives 90\% credible upper limits on $\epsilon$ for the parametric- and cascade-decay benchmarks in the regime $m_X\leq\omega_c/10$.

The present sensitivity is limited mainly by the duration and duty cycle of the search dataset, the usable cat-state size, the directional coverage of a single cavity, counting statistics, and calibration uncertainties. Longer observations would sample the diurnal template more completely and reduce statistical fluctuations; improved state preparation and calibration would reduce the detector-response contribution. Measurements at different locations~\cite{Fukuda:2025zcf,He:2025ovo} or with several modes of one cavity would provide complementary directional responses and an internal test of a candidate modulation. The same data-analysis procedure can also be adapted to searches for other weak vector fields and high-frequency gravitational waves~\cite{Berlin:2021txa,Berlin:2023grv}.

\bigskip



\textit{Acknowledgements.}
This work was supported by the Quantum Science and Technology--National Science and Technology Major Project (Grant No.~2024ZD0302300). J.S. is supported by the National Natural Science Foundation of China under Grant No.~12450006. B.X. is supported by a KIAS Individual Grant No.~PG103301 at Korea Institute for Advanced Study.

\bibliography{refs}

\end{document}